\documentclass[10pt]{article} 

\usepackage[ansinew]{inputenc}
\usepackage{amsmath, amssymb, graphics, amsthm}
\usepackage{epsfig}
\usepackage{color}
\usepackage{undertilde}
\usepackage{fancyhdr} 
\usepackage{bbold}
\usepackage{slashed}

\makeatletter
\@addtoreset{equation}{section}
\makeatother

\newcommand{\be}{\begin{equation}}
\newcommand{\ee}{\end{equation}}
\newcommand{\ba}{\begin{eqnarray}}
\newcommand{\ea}{\end{eqnarray}}

\title{{\sf Towards Loop Quantum Supergravity (LQSG)}\\
{\sf II. p-Form Sector}} 
\author{
{\sf N. Bodendorfer}$^1$\thanks{{\sf 
norbert.bodendorfer@gravity.fau.de}},
{\sf T. Thiemann}$^{1,2}$\thanks{{\sf 
thomas.thiemann@gravity.fau.de,
tthiemann@perimeterinstitute.ca}},
{\sf A. Thurn}$^1$\thanks{{\sf 
andreas.thurn@gravity.fau.de}}\\
\\
{\sf $^1$ Inst. for Theoretical Physics III, FAU Erlangen -- N\"urnberg,}\\
{\sf Staudtstr. 7, 91058 Erlangen, Germany}\\
\\
{\sf and}\\
\\
{\sf $^2$ Perimeter Institute for Theoretical Physics,}\\ 
{\sf 31 Caroline Street N, Waterloo, ON N2L 2Y5, Canada}
}
\date{{\small\sf \today}}

\begin{document} 

\maketitle

{\sf

\begin{abstract}
In our companion paper, we focussed on the quantisation of the Rarita-Schwinger sector 
of Supergravity theories in various dimensions by using an extension of Loop Quantum Gravity to all spacetime dimensions. In this paper, 
we extend this analysis by considering the quantisation of additional bosonic fields necessary 
to obtain a complete SUSY multiplet next to graviton and gravitino in various dimensions. 
As a generic example, we study concretely the quantisation of the $3$-index photon of minimal 
$11d$ SUGRA, but our methods easily extend to more general $p$-form fields. 

Due to the presence of a Chern-Simons term for the $3$-index photon, which is due to local SUSY,
the theory is self-interacting and its quantisation far from straightforward. Nevertheless,
we show that a reduced phase space quantisation with respect to the  $3$-index photon 
Gau{\ss} constraint is possible. Specifically, the Weyl algebra of observables, which 
deviates from the usual CCR Weyl algebras by an interesting twist contribution proportional
to the level of the Chern-Simons theory, admits a background independent state of the 
Narnhofer-Thirring type.
\end{abstract}

}

\newpage

\section{Introduction}
\label{s1}

In our companion papers \cite{BTTI,BTTII,BTTIII,BTTIV,BTTV}, we studied the canonical formulation of 
General Relativity (gravitons) coupled to standard matter in terms of connection variables 
for a compact gauge group without second class constraints in order that 
Loop Quantum Gravity (LQG) quantisation methods, so far formulated only in three and four 
spacetime dimensions \cite{RovelliQuantumGravity, ThiemannModernCanonicalQuantum}, apply. 

The actual motivation for doing this comes from Supergravity and String theory
\cite{GreenBook1, GreenBook2}: String theory is considered as a candidate for a UV completion of 
General Relativity, which in its present formulation requires extra dimensions and supersymmetry. Supergravity
is considered as the low energy effective field theory limit of String theory. One may therefore call String theory a 
top -- bottom approach.  In this series of papers we take first steps towards a 
bottom -- top approach in that we try to canonically quantise the Supergravity theories by 
LQG methods. While String theory in its present form needs a background dependent
and perturbative quantum formulation, the LQG quantum formulation is by design    
background independent and non perturbative. On the other hand, quantum String theory
is much richer above the low energy field theory limit, containing an infinite tower of 
higher excitation modes of the string, which come into play only when approaching the Planck scale and which are necessary in order to find a theory which is finite at least order by 
order in perturbation theory. 

The quantisation of Supergravity
is therefore the ideal arena in which to compare these two complementary approaches
to quantum gravity, which was not possible so far. At least at low energies, that is,
in the semiclassical limit, the two theories should agree with each other, as otherwise 
they would quantise two different classical theories. Evidently, this opens the very
exciting possibility of cross fertilisation between the two approaches, which we are going 
to address in future publications.

The new field content of Supergravity theories as compared to standard matter 
Lagrangians are 1. Majorana (or Majorana-Weyl) spinor fields of spin $1/2,3/2$ 
including the Rarita-Schwinger field (gravitino) and 2. additional bosonic fields
that appear in order to obtain a complete supersymmetry multiplet in the dimension
and the amount $\cal N$ of supersymmetry charges under consideration. The 
treatment of the Rarita-Schwinger sector and its embedding in the framework 
of \cite{BTTI,BTTII,BTTIII,BTTIV,BTTV} was accomplished in \cite{BTTVI}. In this paper, we complete the quantisation
of the extra matter content of many Supergravity theories by considering the quantisation
of the additional bosonic fields, in particular, $p$-form fields. Specifically, 
for reasons of concreteness, we quantise the 
3-index photon of $11d$ Supergravity but it will transpire that the methods employed 
generalise to arbitrary $p$. 

What makes the quantisation possible is that the Gau{\ss} constraints of the 3-index photon 
form an Abelian ideal in the constraint algebra. If this ideal (or subalgebra) would be 
non -- Abelian, then our methods would be insufficient and we most probably would have to use methods from higher gauge theory \cite{BaezAnInvitationTo, BaezInfiniteDImensionalRepresentations,Crane2CategorialPoincare, CraneMeasurableCategoriesAnd, YetterMeasurableCategories}
such as $p$-groups, $p$-holonomies etc., a subject which at the moment is not yet sufficiently
developed from the mathematical perspective (see \cite{BaezHigherGaugeTheory} for the 
state of the art of the subject).
Despite the Abelian character of this additional Gau{\ss} constraint, the quantisation of the 
theory is not straightforward and cannot be performed in complete analogy to the 
treatment of the Abelian Gau{\ss} constraint of standard 1-form matter \cite{ThiemannQSD5}. 
This is due to a Chern-Simons term in the Supergravity action, whose presence is dictated by supersymmetry
and which makes the theory in fact self-interacting, that is, the Hamiltonian is 
a fourth order polynomial in the 3-connection and its conjugate momentum just like in 
Yang-Mills theory. In particular, while one can define a holonomy flux
algebra  as for Abelian Maxwell-theory, the Ashtekar-Isham-Lewandowski representation   
\cite{AshtekarRepresentationsOfThe, AshtekarRepresentationTheoryOf} is inadequate because the Abelian gauge group does not preserve the holonomy flux
algebra.

A solution to the problem lies in performing a reduced phase space quantisation
in terms of a twisted holonomy flux algebra, which is in fact Gau{\ss} invariant. 
We were not able to find a background independent representation of the corresponding Heisenberg algebra, which also 
differs by a twist from the usual one, however, one succeeds when formulating
 the quantum theory 
in terms of the corresponding Weyl elements. The resulting Weyl algebra is not 
of standard form and to the best of our knowledge it has not been quantised before. 
We show that it admits a state of the Narnhofer-Thirring type \cite{NarnhoferCovariantQEDWithout}
whence the Hilbert space representation follows by the GNS construction. 
The Hamiltonian (constraint) can be straightforwardly expressed in terms of the Weyl
elements, in fact it is quadratic in terms of the classical observables, that is, the generators of the 
Heisenberg algebra.\\
\\
This paper's architecture is as follows:\\
\\
In section 2, we sketch the Hamiltonian analysis of the 3-index photon in a self-contained fashion
for the benefit of the reader and in order to settle our notation.  We also describe in detail 
why one cannot straightforwardly apply methods from LQG as mentioned above.\\
In section 3, we display the reduced phase space quantisation solution in terms of the 
twisted holonomy flux algebra. \\
Finally, in section 4, we summarise and conclude.

\section{Classical Hamiltonian Analysis of the 3-Index-Photon Action}
\label{s2}

The Hamiltonian analysis of the full $11d$ SUGRA Lagrangian has been performed in \cite{DiazHamiltonianFormulationOf}. We will review the analysis of the contribution of the 3-index-photon 3-form $A_{\mu\nu\rho}=A_{[\mu\nu\rho]}$ to the $11d$ SUGRA Lagrangian with Chern-Simons term. This part of the Lagrangian is given up to a numerical constant by
\be \label{2.1}
{\cal L}_{{\rm C}}=
-\frac{1}{2}|g|^{1/2} F_{\mu_1..\mu_4}\; F^{\mu_1..\mu_4}
-\alpha |g|^{1/2} F_{\mu_1..\mu_4}\; J^{\mu_1..\mu_4}
-\frac{c}{2}|g|^{1/2} F_{\mu_1..\mu_4}\; F_{\nu_1..\nu_4}\;A_{\rho_1..\rho_3}
\epsilon^{\mu_1..\mu_4\nu_1..\nu_4\rho_1..\rho3} \text{.}
\ee
Here, $F=dA,\;F_{\mu_1..\mu_4}=\partial_{[\mu_1} A_{\mu_2..\mu_4]}$ is the curvature of the 3-index-photon and indices are moved with 
the spacetime metric $g^{\mu\nu}$. Furthermore, $J$ is a totally skew tensor current bilinear 
in the graviton field not containing derivatives, whose explicit form does not need to concern us here, except that it does 
not depend on any other fields. Finally,
$c,\alpha$ are positive numerical constants whose value is fixed by the requirement of 
local supersymmetry \cite{NieuwenhuizenSupergravity}. The number $c$ could be called the level
of the Chern-Simons theory in analogy to $d = D+1=3$.

We proceed to the $10+1$ split of this Lagrangian in a coordinate system with coordinates 
$t,x^a;\;a=1,..,10$ adapted to a foliation of the spacetime manifold. The result of a tedious 
calculation is given by
\ba \label{2.2}
F_{\mu_1..\mu_4}\; F^{\mu_1..\mu_4} &=&
4 F_{ta_1..a_3}\; F^{t a_1..a_3}+F_{a_1..a_4} F^{a_1..a_4} \text{,}
\nonumber\\
F_{ta_1..a_3}\; F^{t a_1..a_3}
&=& G^{a_1..a_3,b_1..b_3} \; F_{ta_1..a_3} \; F_{t b_1..b_3}
-M^{a_1..a_3,b_1..b_4} \; F_{ta_1..a_3} \; F_{b_1..b_4} \text{,}
\nonumber\\
G^{a_1..a_3,b_1..b_3} &=& g^{tt} g^{a_1b_1} g^{a_2 b_2} g^{a_3 b_3}-3g^{ta_1} g^{tb_1}
g^{a_2 b_2} g^{a_3 b_3} \text{,}
\nonumber\\
M^{a_1..a_3,b_1..b_4} &=& g^{a_1 b_2} g^{a_2 b_2} g^{a_3 b_3} g^{tb_4} \text{,}
\nonumber\\
F_{a_1..a_4}\; F^{a_1..a_4} &=& V_1-4M^{a_1..a_3,b_1..b_4} F_{ta_1..a_3} F_{b_1..b_4} \text{,}
\nonumber\\
V_1 &=& g^{a_1b_1} .. g^{a_4 b_4}\; F_{a_1..a_4}\; F_{b_1 .. b_4} \text{,}
\nonumber\\
F_{\mu_1..\mu_4}\; F_{\nu_1..\nu_4}\;A_{\rho_1..\rho_3}
\epsilon^{\mu_1..\mu_4\nu_1..\nu_4\rho_1..\rho_3} &=&
8\epsilon^{a_1..a_3 b_1 .. b_4 c_1..c_3} F_{ta_1.. a_3} F_{b_1..b_4} A_{c_1 .. c_3}
\nonumber\\
&& +3\epsilon^{a_1..a_4 b_1 .. b_4 c_1c_2} F_{a_1.. a_4} F_{b_1..b_4} A_{t c_1 c_2} \text{,}
\nonumber\\ 
J^{\mu_1..\mu_4} F_{\mu_1.. \mu_4} &=& 4 j^{a_1 .. a_3} F_{ta_1 .. a_3}+V_2 \text{,}
\nonumber\\
V_2 &=& J^{a_1.. a_4} F_{a_1 .. a_4} \text{,}
\ea
where we used $\epsilon^{a_1..a_D}=\epsilon^{t a_1 .. a_D}$ and defined  $j^{a_1..a_3}:=J^{t a_1..a_3}$. The potential terms $V_1,V_2$
only depend  on the spatial components of the curvature and do not contain time 
derivatives. 

Using 
\be \label{2.3}
F_{ta_1..a_3}=\frac{1}{4}[\dot{A}_{a_1..a_3}-3\partial_{[a_1} A_{a_2 a_3] t}] \text{,}
\ee
we may perform the Legendre transform. The momentum conjugate to $A$ reads
\ba \label{2.4}
\pi^{a_1..a_3}&=&\frac{\partial {\cal L}}{\partial \dot{A}_{a_1..a_3}} \\
&=&-|g|^{1/2}[G^{a_1..a_3,b_1..b_3} F_{tb_1..b_3} -M^{a_1..a_3,b_1..b_4} F_{b_1.. b_4}
+\alpha j^{a_1.. a_3}] -c \; \epsilon^{a_1..a_3 b_1 .. b_4 c_1 .. c_3} F_{b_1 .. b_4} A_{c_1.. c_3} \text{.} \nonumber
\ea
We may solve (\ref{2.4}) for $F_{t a_1 a_2 a_3}$
\ba \label{2.5}
F_{t a_1 .. a_3} &=& -|g|^{1/2} G_{a_1 .. a_3, b_1.. b_3} \; [\pi^{b_1 .. b_3}+B^{b_1.. b_3} 
+\alpha |g|^{1/2} j^{b_1 .. b_3}] \text{,}
\nonumber\\
B^{a_1..a_3} &=& 
c \epsilon^{a_1..a_3 b_1.. b_4 c_1.. c_3} F_{b_1..b_4} A_{c_1 .. c_3}   
-|g|^{1/2}\; M^{a_1..a_3, b_1.. b_4} F_{b_1.. b_4} \text{,}
\ea
where 
\be \label{2.6}
G_{a_1..a_3, c_1..c_3} G^{c_1.. c_3, b_1.. b_3}=\delta_{[a_1}^{b_1} \delta_{a_2}^{b_2}
\delta_{a_3]}^{b_3}
\ee
defines the inverse of $G$. \\
Inverting (\ref{2.3}) for $\dot{A}$ and using (\ref{2.4}) and  (\ref{2.2}) we obtain for the 
Hamiltonian after a longer calculation
\ba \label{2.7}
H &=& \int\;d^{10}x\; \left\{ \dot{A}_{a_1 .. a_3} \pi^{a_1 .. a_3} - {\cal L} \right\}
\nonumber\\
&=& -\int\; d^{10}x \Big\{ 3 A_{ta_1 a_2}  \; G_C^{a_1 a_2}+2|g|^{-1/2} 
G_{a_1..a_3,b_1..b_3} [\pi+B+\alpha j]^{a_1..a_3} \; [\pi+B+\alpha j]^{b_1.. b_3} \nonumber \\
& & ~~~~~~~~~~~~~~ +|g|^{1/2}[V_1/2+\alpha V_2] \Big\} \text{,}
\nonumber\\
G_C^{a_1 a_2} &:=& \partial_{a_3}\pi^{a_1..a_3}
-\frac{c}{2}\epsilon^{a_1 a_2 b_1..b_4 c_1..c_4} F_{b_1..b_4} F_{c_1..c_4} \text{,}
\ea
where an integration by parts has been performed in order to isolate the Lagrange 
multiplier $A_{ta_1 a_2}$. Using the ADM frame metric components 
\be \label{2.8a}
g^{tt}=-1/N^2,\;g^{ta}=N^a/N^2,\;g^{ab}=q^{ab}-N^a N^b/N^2;\;\;
g_{tt}=-N^2+q_{ab} N^a N^b,\; g_{ta}=q_{ab} N^b,\;g_{ab}=q_{ab} \text{,}
\ee
with $q_{ab}$ the induced metric on the spatial slices and lapse respectively
shift functions $N,N^a$ we can easily decompose the piece of $H$ independent
of the 3-index Gau{\ss} constraint $G_C^{a_1 a_2}$ into the contributions 
$N^a {\cal H}_{Ca} +N {\cal H}_C$ to the spatial diffeomorphism constraint and 
Hamiltonian constraint, however, we will not need this at this point.\\
We will drop the subscript $C$ in what follows, since in this paper we are only interested in 
the $p$-form sector. We smear the Gau{\ss} constraint with a 2-form $\Lambda$, that is
\be \label{2.8b}
G[\Lambda]:=\int\; d^{10}x\; \Lambda_{ab} \; G^{ab}
\ee
and study the gauge transformation behaviour of the canonical 
pair $(A_{abc},\pi^{abc})$ with non-vanishing Poisson brackets
\be \label{2.9}
\{\pi^{a_1..a_3}(x),A_{b_1..b_3}(y)\}=\delta^{(10)}(x,y)\; \delta^{a_1}_{[b_1} \delta^{a_2}_{b_2} \delta^{a_3}_{b_3]} \text{.}
\ee
We find 
\ba \label{2.10}
\{G[\Lambda],A_{a_1.. a_3}\} &=& -\partial_{[a_1} \Lambda_{a_2 a_3]} \text{,}
\nonumber\\
\{G[\Lambda],\pi^{a_1.. a_3}\} &=& c\; \epsilon^{a_1..a_3 b_1..b_3 c_1 .. c_4}
\partial_{[b_1} \Lambda_{b_2 b_3]} F_{c_1..c_4}\text{.}
\ea
These equations can be written more compactly in differential form language, in terms 
of which they are easier to memorise. Introducing the dual 7-pseudo-form\footnote{Notice
that $\pi^{abc}$ is a tensor density of type $T^3_0$ and density weight one.} 
 \be \label{2.11}
 (\ast\pi)_{a_1..a_7}:=\frac{1}{3!\;7!} \epsilon_{b_1..b_3 a_1..a_7} \pi^{b_1..b_3}
 \ee
 we may write (\ref{2.10}) as 
 \be \label{2.12}
 \delta_\Lambda A=-d\Lambda,\;\;\;\;
 \delta_\Lambda\; \ast\pi=c\; (d\Lambda)\wedge F \text{.}
 \ee
 Since the right hand side of (\ref{2.12}) is closed, in fact exact, it would seem that
 the observables of the theory can be coordinatised by integrals of $A$ and $\ast \pi$
 respectively over closed 3-submanifolds or 7-submanifolds respectively. 
 
 The $G(\Lambda)$ generate an Abelian ideal in the constraint algebra since
 \be \label{2.12a}
 \left\{G[\Lambda],G[\Lambda'] \right\}=0,\;\;\;\; \left\{G[\Lambda],H(x) \right\}=0 \text{,}
 \ee
 where $H(x)$ is the integrand of $H$ in (\ref{2.7}) and since the only $\pi$ or $A$
 dependent contributions to the Hamiltonian and spatial diffeomorphism constraints are 
 contained in $H(x)$.
 
 We see that due to the non vanishing Chern-Simons constant $c$, the transformation 
 behaviour of $\ast \pi$ differs from the transformation behaviour with respect to 
 the higher dimensional analog of the usual Maxwell type of Gau{\ss} law, which would be just the divergence term
 $\partial_{a_1} \pi^{a_1..a_3}$. In particular, $\pi^{abc}$ itself is {\it not} gauge invariant.
 This ``twisted'' Gau{\ss} constraint (\ref{2.7}) can be  written in the form
 \be \label{2.13}
 G^{a_1 a_2} := \partial_{a_3}[\pi^{a_1..a_3}
-\frac{c}{2}\epsilon^{a_1 ..a_3 b_1..b_3 c_1..c_4} A_{b_1..b_3} F_{c_1..c_4} ]  
=:\partial_{a_3} \pi^{\prime a_1.. a_3} \text{,}
\ee
which suggests to introduce a new momentum $\pi'$. Unfortunately, this does not work
because $\ast(\pi'-\pi)=A\wedge F$ does not have a generating functional $K$ 
with $\delta K/\delta A=A\wedge F$, since the only possible candidate 
$K=\int A\wedge A\wedge F\equiv 0$ identically vanishes in the dimensions considered
here.
Since this is not the case, the Poisson brackets of $\pi'$ with itself do not vanish
and neither is $\pi'$ gauge invariant as we will see below, so that there is no advantage 
of working with $\pi'$ as compared to $\pi$.

The presence of the twist term in the Gau{\ss} constraint leads to the following difficulty
when trying to quantise the theory on the usual LQG type kinematical Hilbert space:\\
Such a Hilbert space would roughly be generated by a holonomy flux algebra 
constructed from holonomies 
\be \label{2.14}
A(e)=\exp(i\int_e\; A),\;\;\;\; \pi(S)=\int_S\; \ast\pi \text{,}
\ee
where $e$ and $S$ are oriented 3-dimensional and 7-dimensional submanifolds respectively,
which we call ``edges'' and surfaces in what follows.
One could then study the GNS Hilbert space representation generated by the LQG
type of positive linear functional
\be \label{2.15}
\omega(f \pi(S_1)..\pi(S_n))=0,\;\;\;\; \omega(f)=\mu[f] \text{,}
\ee
where $\mu$ is an LQG type measure on a space of generalised connections
$\overline{{\cal A}}$. One can define 
it abstractly by requiring that the charge network functions 
\be \label{2.15a}
T_{\gamma,n}=\prod_{e\in \gamma}\; A(e)^{n_e},\;n_e\in \mathbb{Z}
\ee 
form an orthonormal basis in the corresponding ${\cal H}=L_2(\overline{{\cal A}},\mu)$,
see \cite{ThiemannModernCanonicalQuantum} for details. Here, a graph $\gamma$ is a collection of edges which are disjoint up
to intersections in ``vertices'', which are oriented 2-manifolds. The possible intersection 
structure of these cobordisms should be tamed by requiring that all submanifolds are semi-analytic.\\
Up to here everything is in full analogy with LQG. The problem is now to isolate 
the Gau{\ss} invariant subspace of the Hilbert space: While the connection transforms 
as in a theory with untwisted Gau{\ss} constraint, it appears that we can solve it by requiring
 that charges add up to zero at vertices. However, this does not work because while such 
 a vector is annihilated by the divergence term in $G^{ab}$, it is not by the second 
 term $\propto A\wedge F$. Even more disastrous, the term $A\wedge F$ does not 
 exist in this representation which is strongly discontinuous in the holonomies so that operators 
 $A,F$ do not exist. Finally, although $\pi$ is not Gau{\ss} invariant, it leaves this would 
 be gauge invariant subspace invariant, which reveals that this subspace is not 
 the kernel of the twisted Gau{\ss} constraint. 
 
 We therefore must be more sophisticated. Since the $A$ dependent terms in $G$ cannot be
 quantised on the kinematical Hilbert space, we must exponentiate it:\\
 Consider the Hamiltonian flow of $G[\Lambda]$
 \be \label{2.16}
 \exp(\{G[\Lambda],\cdot \}) A=A-d\Lambda,\;\;\;\;
  \exp(\{G[\Lambda],\cdot \}) \ast\pi=\ast\pi+c (d\Lambda)\wedge F \text{,}
 \ee
 which is a Poisson automorphism $\alpha_\Lambda$ (canonical transformation) and one would like 
 to secure that an implementation of the corresponding automorphism group 
 $\alpha_\Lambda\circ \alpha_{\Lambda'}=\alpha_{\Lambda+\Lambda'}$ 
 by unitary operators $U(\Lambda)$ exists.  The $U(\Lambda)$ would correspond to the desired exponentiation
 of the Gau{\ss} constraint. One way of securing this is  by looking for 
 an invariant state $\omega=\omega\circ \alpha_\Lambda$ on the holonomy - flux algebra
 (see \cite{HaagLocalQuantumPhysics}) for the details for this construction). This would then open the possibility
 that the Gau{\ss} constraint can be solved by group averaging methods.
 The first problem is that the automorphisms do not preserve the holonomy flux 
 algebra because there appears an $F$ on the right hand side of (\ref{2.16}) which should appear
 exponentiated  in order that the algebra closes. This forces us 
 to pass to exponentiated fluxes, that is, to the corresponding Weyl algebra defined by
 exponentials of $\pi, A$. This algebra is now preserved by the automorphisms, as one can 
 see by an appeal to the Baker-Campbell-Hausdorff formula.
 However, we now see that the state (\ref{2.15}) is not invariant, because 
 \be \label{2.17}
 \omega(e^{i\pi(S)})=1,\;\;\;\;
 \omega(\alpha_\Lambda(e^{i\pi(S)}))=
 \omega(e^{i[\pi(S)+c\int_S d\Lambda \wedge F]})=0
 \ee
 for suitable choices of $\Lambda$. 
In the GNS Hilbert space we would like 
to have unitary operators $U(\Lambda)$
such that for any element $W$ in the Weyl
algebra we have $U(\Lambda)\pi(W) U(\Lambda)^\ast
=\pi(\alpha_\Lambda(W))$. Then (\ref{2.17}) is compatible
with unitarity only if the LQG vacuum $\Omega$ is not
invariant under $U(\Lambda)$.   
Now the operator $U(\Lambda)$ should
correspond to $\exp(iG[\Lambda])$ and using
a calculation similar to (\ref{2.12a}) and the BCH
formula one shows that
on the LQG vacuum $\Omega=1$ it reduces formally to 
$U(\Lambda)\Omega=\exp(ic/2\int \Lambda\wedge F\wedge F)\Omega$
which is ill defined as it stands. We must therefore
{\it define} $U(\Lambda)\Omega$ to be some state
in the GNS Hilbert space which has a component 
orthogonal to the vacuum and such that the representation
property $U(\Lambda) U(\Lambda')=U(\Lambda+\Lambda'),\;\;
U(\Lambda)^\ast=U(-\Lambda)$  (possibly up to a projective
twist) holds. We did not succeed to find a solution to
this problem indicating that a unitary implementation
of the Gauss constraint is impossible in the LQG representation
and even it were possible, the strategy outlined in the next section
is certainly more natural.   
We also remark that solving the constraint by group averaging methods becomes non trivial if not impossible in case of the non existence of $U(\Lambda)$. Even if we could somehow construct the Gau{\ss} invariant Hilbert space, the observables 
 $A(e),\exp(i\pi(S))$ with $\partial e=\partial S=\emptyset$, which leave the physical Hilbert space 
 invariant, are insufficient to approximate (for small $e,S$) the $\pi$ dependent terms appearing in the Hamiltonian (\ref{2.7}), as one can check explicitly.
 
  \section{Reduced Phase Space Quantisation}
 \label{s3} 
 
 In the previous section, we established that a quantisation in strict analogy 
 to the procedure followed in LQG does not work. While a rigorous kinematical Hilbert
 space can be constructed, the Dirac operator constraint method of looking 
 for the kernel of the Gau{\ss} constraint is problematic. As an alternative, a reduced phase 
 space quantisation suggests itself. This has a chance to work due to the 
 observation (\ref{2.12a}) which demonstrates that $H(x)$ only depends on observables.
 Indeed, $H(x)$ depends, except for $G^{ab}$ which is a trivial observable since it is 
 constrained to vanish, only on the combination $\pi+B+\alpha j$. Obviously $j$ trivially
 Poisson commutes with $G$. Unpacking $B$ from (\ref{2.5}), we see that 
 $\pi+B$ is a linear combination (with only metric dependent coefficients) of $F$ and 
 \be \label{3.1}
 P^{abc}:=\pi^{abc}+
 c \epsilon^{abc d_1.. d_4 e_1.. e_3} F_{d_1..d_4} A_{e_1 .. e_3}
 \;\; \Leftrightarrow\;\; \ast P=\ast\pi+c\; A\wedge F   \text{,}
\ee
which suggests that $\{G(\Lambda),P^{abc}(x)\}=0$ because $F$ is already invariant.
This indeed can be verified
using (\ref{2.12})
\be \label{3.2}
\delta_\Lambda \ast P=\delta_\Lambda \ast\pi+c \delta_\Lambda A\wedge F
=0 \text{.}
\ee
Our classical observables therefore are coordinatised by  the 4-form and 7-form 
$F=dA$ and $\ast P=\ast\pi +c A\wedge F$ respectively. Since $F$ is exact,
it is determined entirely by a 3-form modulo an exact form, which in turn 
is parametrised by a 2-form. This 2-form worth of gauge freedom matches the number 
of Gau{\ss} constraints which can be read as a condition on $\pi$. Thus, on the constraint
surface, the number of degrees of freedom contained in $F$ and $P$ match. 
 
 We compute the observable algebra. Let $f$ be a 3-form and $h$ 
 a 6-form with dual $\ast h$ (a totally skew 
 4-times contravariant tensor pseudo density) and smear the observables
 with these
 \be \label{3.3}
 P[f]:=\int \; d^{10}x\; f_{a_1..a_3}\; P^{a_1..a_3}=\int f\wedge \ast P ,\;\;\;\;
 F[h]:=\int\; d^{10}x\; \; (\ast h)^{a_1..a_4} F_{a_1..a_4}=\int\; h\wedge F \text{.}
 \ee
 Then, we find after a short computation 
 \be \label{3.4}
 \{F[h],F[h']\}=0,\;\;\;\;\{P[f],F[h]\}=\int\; h\wedge df,\;\;\;\;
 \{P[f],P[f']\}=-3 c \; F[f \wedge f'] \text{.}
 \ee
 Thus, the observable algebra closes but $P$ is not conjugate to $F$. 
 
 The form of the observable algebra (\ref{3.4}) reveals the following:\\
 Typically, background independent representations tend to be discontinuous 
 in at least one of the configuration or the momentum variable. For instance, in LQG
 electric fluxes exist in non exponentiated form, but connections do not. Let us assume that 
 we find such a representation in which $F[h]$ does not exist so that we have to consider 
 instead its exponential (Weyl element). Then (\ref{3.4}) tells us that in such a representation 
 automatically also $P[f]$ cannot be defined, because if it could, then its commutator would
 exist, which however is proportional to some $F$ which is a contradiction. Hence, either 
 both $F,P$ exist or only both of their corresponding Weyl elements. 
 
 We did not manage to find a representation in which the Weyl elements 
 \be \label{3.5}
 W[h,f]:=\exp \left(i(F[h]+P[f]) \right)
 \ee
 are strongly continuous operators in both $f,h$. However, we did find one in which 
 they are discontinuous in both $h,f$. This representation was studied in the context
 of QED in \cite{NarnhoferCovariantQEDWithout} and was applied to an LQG type of quantisation of the 
 closed bosonic string in \cite{ThiemannTheLQGString1}. Before we define it, we must first 
 define the Weyl algebra generated by the Weyl elements (\ref{3.5}).
 The $^\ast$-relations are obvious,
 \be \label{3.6}
 W[h,f]^\ast=W[-h,-f] \text{.}
 \ee
 However, the product relations are very interesting and non trivial, because they require the 
 generalisation of the Baker-Campbell-Hausdorff formula \cite{CampbellBCH1, CampbellBCH2, BakerBCH1, BakerBCH2, BakerBCH3, HausdorffBCH} to higher commutators \cite{MagnusOnTheExponential}.
 Suppose that $X,Y$ are operators on some Hilbert space such that the triple commutators 
 $[X,[X,Y]]$ and $[Y,[Y,X]]$ commute with both $X$ and $Y$. This formally applies to our case 
with $X=F[h]+P[f],\;\;Y=F[h']+P[f']$, which obey the canonical commutation relations
(we set $\hbar=1$ for simplicity)
\be \label{3.7}
[X,Y]:=i \{X,Y\}=i \left\{[\int(h'\wedge df-h \wedge df')] \;\mathbb{1}-3 c F[f\wedge f'] \right\} \text{.}
\ee
From this follows for the triple commutators
\ba \label{3.8}
~[X,[X,Y]] &=& -3c (i)^2\;\{P[f],F[f\wedge f']\}=3 c \int\;f\wedge f'\wedge df\;\mathbb{1} \text{,}
\nonumber\\
~[Y,[Y,X]] &=& 3c (i)^2\;\{P[f'],F[f\wedge f']\}=-3 c \int\;f\wedge f'\wedge df'\;\mathbb{1} \text{,}
\ea
which thus are in the centre of the algebra.

The BCH formula for the case of all triple commutators commuting with $X,Y$ reads
\be \label{3.9}
e^X\; e^Y=e^{X+Y+\frac{1}{2}[X,Y]+\frac{1}{12}([X,[X,Y]]+[Y,[Y,X]])} \text{,}
\ee
which can also be proved using elementary methods.
From this it is easy to derive the also useful Zassenhaus formula \cite{MagnusOnTheExponential}
\be \label{3.10}
e^{X+Y}=e^{X}\; e^Y\; e^{-\frac{1}{2}[X,Y]}\;e^{-\frac{1}{6}([X,[X,Y]]+2[Y,[X,Y]])} \text{.}
\ee
Putting all these together, we obtain the Weyl relations
\be \label{3.11}
W[h,f]\;W[h',f']=W[h+h'+\frac{3 c}{2}f\wedge f',f+f']\;\;
\exp \left(\frac{i}{4}\int \left[2(h\wedge df'-h'\wedge df)-cf\wedge f'\wedge d(f-f') \right] \right) \text{.}
\ee
Hence also the Weyl relations get twisted as compared to the situation with $c=0$. 
Notice that the first term in the phase is antisymmetric under the exchange 
$(h,f)\leftrightarrow (h',f')$, while the second is symmetric.

In order to obtain a representation of this $^\ast$-algebra $\mathfrak{A}$ generated by the Weyl elements, it is sufficient to find a 
positive linear functional. We consider the Narnhofer-Thirring type of functional
\be \label{3.12}
\omega(W(h,f))=\left\{ \begin{array}{cc} 
1 & h=f=0\\
0 & {\rm else}
\end{array}  \right.
\ee
and show that it is positive definite on $\mathfrak{A}$. Let 
\be \label{3.13}
a:=\sum_{k=1}^N\; c_k\; W[z_k] 
\ee
be a general element in $\mathfrak{A}$, where $N\in \mathbb{N}$, $c_k\in \mathbb{C}$ and the 
$z_k=(h_k,f_k)$ are arbitrary, where without loss of generality 
$z_k\not=z_l$ for $k\not=l$. We have 
\ba \label{3.14}
\omega(a^\ast a) 
&=&
\sum_{k,l=1}^N \; \bar{c}_k\; c_l\; \omega(W[-z_k]\; W[z_l])
\nonumber\\
&=&
\sum_{k,l=1}^N \; \bar{c}_k\; c_l\; \omega(W[z_{kl}])\exp(i\alpha_{kl}) \text{,}
\nonumber\\
z_{kl} &=& (-h_k+h_l-\frac{3}{2}c f_k\wedge f_l,\;\;-f_k+f_l) \text{,}
\nonumber\\
\alpha_{kl} &=& 
\frac{1}{4}\int[2(-h_k\wedge df_l+h_l\wedge df_k)-cf_k \wedge f_l\wedge d(f_k+f_l)] \text{.}
\ea
For $k=l$, we have $z_{kl}=\alpha_{kl}=0$ because $f_k,f_l$ are 3-forms. For $k\not=l$,
we must have either $f_k\not=f_l$ or $h_k\not=h_l$ or both. If $f_k\not=f_l$, then 
obviously $z_{kl}\not= 0$. If $f_k=f_l$, then necessarily $h_k\not= h_l$ and 
$z_{kl}=(-h_k+h_l,0)\not=0$. By definition (\ref{3.12}) then
\be \label{3.15}
\omega(a^\ast a)=\sum_{k=1}^N \; |c_k|^2 \ge 0;\;\; \omega(a^\ast a)=0\;\;\Leftrightarrow\;\;
a=0
\ee
is positive definite. Thus, the left ideal $\mathfrak{I}=\{a\in\mathfrak{A};\;\;
\omega(a^\ast a)=0\}=\{0\}$ is trivial and the Hilbert space representation is given by the 
GNS data \cite{HaagLocalQuantumPhysics}:\\
The cyclic vector is $\Omega=\mathbb{1}$, the Hilbert space ${\cal H}$ 
is the Cauchy completion of $\mathfrak{A}$ in the scalar product 
$<a,b>:=\omega(a^\ast b)$ and  the representation is simply $\pi(a) b:=ab$
on the common dense domain ${\cal D}=\mathfrak{A}$. \\
The representation is evidently strongly discontinuous in both $h,f$ and while 
cyclic, it is not irreducible.  Equivalently, $\omega$ is not a pure state \cite{BratteliOperatorAlgebrasAnd1, BratteliOperatorAlgebrasAnd2}.\\
\\
The question left open to answer is whether the algebra and the state $\omega$ are still 
well defined when restricting the smearing functions $(h,f)$ to the form factors of 
4-surfaces and 7-surfaces respectively. The bearing of this question is that in the 
Hamiltonian constraint the functions $F$ and $\ast P$ appear in such a way, that in a 
discretisation of it, which results from replacing the integral by Riemann sums in the 
spirit of \cite{ThiemannQSD1}, these functions are naturally smeared over 4-surfaces and 7-surfaces 
respectively. They could thus be approximated by Weyl elements.\\
To answer this question, let $S_4,S_7$ be general 4 and 7 surfaces respectively. Consider the 
distributional forms (``form factors'') 
\ba \label{3.16} 
h^{S_4}_{a_1..a_6}(x) &:=& 
\int_{S_4} \epsilon_{a_1..a_6 b_1..b_4} dy^{b_1}\wedge dy^{b_4}\;
\delta(x,y) \text{,}
\nonumber\\
f^{S_7}_{a_1..a_3} (x) & :=& \int_{S_7} \epsilon_{a_1..a_3 b_1..b_7} dy^{b_1}\wedge dy^{b_7}\;
\delta(x,y) \text{.}
\ea
Then 
\be \label{3.17}
F[h^{S_4}]=\int_{S_4} F,\;\;\;\; P[f^{S_7}]=\int_{S_7}\; \ast P \text{.}
\ee

Thus, the natural integrals of $F,P$ over surfaces can be reexpressed in terms of 
distributional 6 forms and 4-forms respectively.
It remains to check whether the exterior derivative and product combinations of these distributional 
forms appearing in the multiple Poisson brackets of (\ref{3.17}) and in the Weyl
relations remain meaningful. Three types of exterior derivative and product expressions appear.
The first is, using formally Stokes theorem
\be \label{3.18}
\int h^{S_4}\wedge d f^{S_4}=\int_{S_4} df^{S_7}=\int_{\partial S_4} f^{S_7}
=\int_{\partial S_4} dx^{a_1}\wedge..\wedge dx^{a_3}
\epsilon_{a_1..a_3 b_1..b_7}\int_{S_7}\; dy^{b_1}\wedge..\wedge dy^{b_7}
\; \delta(x,y)
=:\sigma(\partial S_4,S_7) \text{.}
\ee
The integral is supported on $\partial S_4\cap S_7$ and we can decompose this set into
components (submanifolds) which are 0,1,2,3-dimensional. The number of these components
will be finite if the surfaces are semianalytic. We define the {\it intersection number}
$\sigma(\partial S_4,S_7)$ to be zero for the 1,2,3-dimensional components and by 
(\ref{3.18}) for the isolated intersection points, which then takes the values $\pm 1$. This can be justified by the same regularisation as in LQG for the holonomy flux algebra \cite {ThiemannModernCanonicalQuantum}. 

The second type of integral is given by $F[f^{S_7}\wedge f^{S_7'}]$. The support 
of the integral will be on $S^{S_7}\cap S^{S_7'}$ and in $D=10$ dimensions this will
decompose into components that are at least 4-dimensional. By the same regularisation 
as in \cite{ThiemannModernCanonicalQuantum}, one can remove the higher dimensional components and thus keep
only the 4-dimensional ones. In what follows, we thus assume that $S_4:=S_7\cap S_7' $
is a single 4-dimensional component, otherwise the non vanishing contributions are 
over a sum of those. We have 
\ba \label{3.19}
F[f^{S_7}\wedge f^{S_7'}] &=& \int_{S_7}\; F \wedge f^{S_7'} 
\\
&=&
\int_{S_7}\; dx^{a_1}\wedge..\wedge dx^{a_4}\wedge dx^{b_1}\wedge..\wedge dx^{b_3}
\epsilon_{b_1..b_3 c_1..c_7}\int_{S_7'}\; dy^{c_1}\wedge..\wedge dy^{c_7}\;
\delta(x,y)\; F_{a_1..a_4}(x) \text{.}
\nonumber
\ea
By assumption, we have embeddings 
\be \label{3.20}
X_{S_7}:\; U\to S_7;\;\;  \;\;
Y_{S'_7}:\; V\to S'_7;\;\;  \;\;
Z_{S_4}:\; W\to S_4,
\ee
with open subsets $U,V$ of $\mathbb{R}^7$ and an open subset $W$ of $\mathbb{R}^4$
respectively, whose coordinates will be denoted by $u,v,w$ respectively. The 
condition $X_{S_7}(u)=Y_{S_7'}(v)=Z_{S_4}(w)$ is solved by solving $u,v$ for $w$, 
which leads to $u=u(w),\; v=v(w)$. Since the integrals are reparametrisation 
invariant, in the neighbourhood of $S_4$ on both $S_7$ and 
$S_7'$ therefore we may use adapted coordinates so that $w^I=u^I=v^I,\;I=1,..,4$
on $S_4$ and $u^I,v^I,\;I=5,..,7$ denote the transversal coordinates, which 
take the value $0$ on $S_4$. In this parametrisation both $U,V$ are of the 
form $U=W\times U',\;V=W\times V'$ for some 3-dimensional subsets $U',V'$ of 
$\mathbb{R}^3$. It follows $Z(w)=X(w,0)=Y(w,0)$ in this parametrisation.
The $\delta$ distribution is then supported on 
$u^I=v^I,\;I=1,..,4$ and $u^I=v^I=0,\;\;I=5,..,7$ and we have in the neighbourhood 
of $S_4$
\be \label{3.22}
X^a(u)-Y^a(v)=-\sum_{I=1}^4 Y^a_I(u,0) \left[u^I-v^I \right]+\sum_{I=5}^7 \left[X^a_I(u,0) u^I-Y^a_I(u,0) v^I \right] \text{.}
\ee
We can now solve the $\delta$ distribution in 
(\ref{3.19}) by performing the integral over $u^5,..,u^7,v^1,..,v^7$ 
and find with the notation $X^a_I=\partial X_{S_7}^a(u)/\partial u^I$ and 
$Y^a_I=\partial Y_{S'_7}^a(v)/\partial v^I$  
etc.
\ba \label{3.21}
F[f^{S_7}\wedge f^{S_7'}] 
&=&
\int_U\; d^7u  \; \epsilon^{I_1..I_7}\; 
\left[X^{a_1}_{I_1}..X^{a_4}_{I_4} X^{b_1}_{I_5}..X^{b_3}_{I_7} \right](u)
\epsilon_{b_1.. b_{10}}\int_V\; d^7 v\;\epsilon^{J_1..J_7}\;
\left[Y^{b_4}_{J_1}..Y^{b_{10}}_{J_7} \right](v) \; \times 
\nonumber\\
&& \delta \left(X(u),Y(v))\; F_{a_1..a_4}(X(u) \right)
\nonumber\\
&=& -
\int_W\; d^4w  \; \epsilon^{I_1..I_7}\; 
\left[Z^{a_1}_{I_1}..Z^{a_4}_{I_4}\right](w)     \;\epsilon^{J_1..J_7}\; F_{a_1..a_4}(Z(w)) \:
\epsilon_{I_5 .. I_7 J_1 .. J_7} \: \times
 \nonumber \\
& &~~~~~~~~~~~~~\; \left[{\rm sgn} \left(\det \left(\frac{\partial(X(u)-Y(v))}{\partial(u^5,..,u^7, v^1,..,v^7)} \right)_{v^I=u^I=w^I;I=1,..,4;v^I=u^I=0; I=5,..,7} \right) \right]\;
\nonumber\\
&&
\nonumber\\
&=:& -3!\; 7! \tilde{\sigma}(S_7,S'_7) F[h^{S_4}] \text{,}
\ea
where the $10d$ antisymmetric symbol is in terms of the coordinates $u^5,..,u^7,v^1,..,v^7$
and in the last step we noticed that the range of $I_1..I_4$ is restricted to $1..4$. 
Also, we assumed that the sign function under the integral is constant and equal
to $\tilde{\sigma}(S_7,S'_7)$ on $S_4$ (which defines this function), otherwise we must decompose $S_4$ further.
Under this assumption, we conclude the form factor identity 
\be \label{3.23}
f_{S_7}\wedge f_{S_7'}=-3!\;7!\;\tilde{\sigma}(S_7,S'_7) h^{S_7\cap S'_7} \text{.}
\ee

Finally, we consider the integral of the third type, 
which now combining (\ref{3.18}) and (\ref{3.24}) is easily calculated 
\be \label{3.24}
\int f_{S_7}\wedge f_{S'_7}\wedge df_{S_7}
=-3! \; 7!\; \tilde{\sigma}(S_7,S'_7)\int h^{S_7\cap S'_7}\wedge df^{S_7}
=-3!\; 7!\; \tilde{\sigma}(S_7,S'_7)\sigma(\partial (S_7\cap S'_7),S_7)=0 \text{,}
\ee
because $\partial (S_7\cap S'_7)\subset S_7$ for which $\sigma$ vanishes by definition.

In order to make this restricted Weyl algebra close, we now have to decide whether 
the form factors should only be added with integer valued coefficients \cite{ThiemannQSD5} 
or with real valued ones \cite{AshtekarPolymerAndFock,KaminskiBackgroundIndependentQuantizations1, KaminskiBackgroundIndependentQuantizations2}.
In the latter case we do not need to do anything and the restricted 
Weyl algebra already closes. In the former case we must replace the form factors 
$f^{S_7}$  by $\frac{1}{\sqrt{3!\;7!\;3c/2}} f^{S_7}$, such that in the simplest situation we have
\be \label{3.25}  
W[S_4,S_7]\;W[S'_4,S'_7]=W[S_4+S'_4-\tilde{\sigma}(S_7,S'_7) S_7\cap S'_7,S_7+S'_7]\;\;
\exp \left(\frac{i}{2} \left[\sigma(\partial S_4,S'_7)-\sigma(\partial S'_4, S_7) \right] \right) \text{,}
\ee
from which the general case can be easily deduced.\\ 
\\
We conclude that the restricted Weyl
algebra is well defined in either case. Thus, wherever $P$ or $F$ appear in the Hamiltonian
constraint, we follow the general regularisation procedure outlined in \cite{ThiemannQSD1}, which  
employs a combination of  spatial diffeomorphism invariance and an infinite refinement limit of a Riemann sum approximation of the Hamiltonian 
constraint in terms of $P[S_7]$ and $F[S_4]=A[\partial S_4]$, which we approximate 
for instance by $\sin(P[S_7]),\;\sin(F[S_4])$ similar as in LQG. The details are obvious and are 
left to the interested reader. 

\section{Conclusions}
\label{s4}

Supergravity theories typically need additional bosonic fields next to the graviton, in order 
to obtain a SUSY multiplet (representation) containing the gravitino. In this paper, we focussed on minimal $11d$ SUGRA for 
reasons of concreteness (and its relevance for lower dimensional SUGRA theories), 
which contains the $3$-index photon in the bosonic sector.
However, our analysis is easily generalised to arbitrary $p$-form fields. Without the 
Chern-Simons term in the action (i.e. $c=0$) the analysis would be straightforward 
and in complete analogy to the background independent treatment of Maxwell theory in
$D+1=4$ dimensions \cite{ThiemannQSD5}. In particular, the Hamiltonian constraint would be quadratic in the 
$3$-form field and its conjugate momentum, which thus would reduce to a free field theory
when switching off gravity. However, with the Chern-Simons term ($c\not=0$) the Hamiltonian 
constraint becomes in fact quartic in the connection and thus becomes self-interacting
even when switching off gravity, just like in non Abelian Yang-Mills theories.

It is therefore the more astonishing that we can quantise the resulting $^\ast$-algebra 
of observables (with respect to the $3$-index-Gau{\ss} constraint) rigorously, even though the 
theory is self-interacting. In fact, in terms of the observables, the Hamiltonian constraint
is a quadratic polynomial, however, the price to pay is that the observable algebra is 
non standard. Yet, the resulting Weyl algebra can be computed in closed form and 
we found at least one non trivial and background independent representation thereof, 
which nicely fits into the background independent quantisation of the gravitational 
degrees of freedom in the contribution to the Hamiltonian constraint depending on the 
$3$-index-photon.

There are many open questions arising from the present study. One of them concerns the 
reducibility of the GNS representation found, which involves a mixed state. It would be 
nice to have control over the superselection sectors of the theory and, in particular,
to analyse whether the cyclic GNS vector is not already cyclic for the Abelian subalgebra
generated by the $W[h,0]$. Next, it is worthwhile to study the question whether this 
algebra admits regular representations for both $P$ and $F$, because then the GNS Hilbert space would admit a measure theoretic interpretation as an $L_2$ space. Finally, it is certainly 
necessary to work out the cobordism theory of relevance when restricting
the Weyl algebra to distributional $4$-form and $7$-form factors as smearing functions which is 
only sketched in this paper. We plan to revisit these questions in future publications.\\
\\
\\
\\
{\bf\large Acknowledgements}\\
NB and AT thank Alexander Stottmeister, Derek Wise, and Antonia Zipfel for numerous discussions and the German National Merit Foundation for financial support. The part of the research performed at the Perimeter Institute for Theoretical Physics was supported in part by funds from the Government of Canada through NSERC and from the Province of Ontario through MEDT.

\newpage

\bibliography{pa93pub.bbl}

\providecommand{\href}[2]{#2}\begingroup\raggedright\begin{thebibliography}{10}

\bibitem{BTTI}
N.~Bodendorfer, T.~Thiemann, and A.~Thurn, ``{New variables for classical and
  quantum gravity in all dimensions: I. Hamiltonian analysis},'' {\em Classical
  and Quantum Gravity} {\bf 30} (2013) 045001, {\tt arXiv:1105.3703 [gr-qc]}.

\bibitem{BTTII}
N.~Bodendorfer, T.~Thiemann, and A.~Thurn, ``{New variables for classical and
  quantum gravity in all dimensions: II. Lagrangian analysis},'' {\em Classical
  and Quantum Gravity} {\bf 30} (2013) 045002, {\tt arXiv:1105.3704 [gr-qc]}.

\bibitem{BTTIII}
N.~Bodendorfer, T.~Thiemann, and A.~Thurn, ``{New variables for classical and
  quantum gravity in all dimensions: III. Quantum theory},'' {\em Classical and
  Quantum Gravity} {\bf 30} (2013) 045003, {\tt arXiv:1105.3705 [gr-qc]}.

\bibitem{BTTIV}
N.~Bodendorfer, T.~Thiemann, and A.~Thurn, ``{New variables for classical and
  quantum gravity in all dimensions: IV. Matter coupling},'' {\em Classical and
  Quantum Gravity} {\bf 30} (2013) 045004, {\tt arXiv:1105.3706 [gr-qc]}.

\bibitem{BTTV}
N.~Bodendorfer, T.~Thiemann, and A.~Thurn, ``{On the implementation of the
  canonical quantum simplicity constraint},'' {\em Classical and Quantum
  Gravity} {\bf 30} (2013) 045005, {\tt arXiv:1105.3708 [gr-qc]}.

\bibitem{RovelliQuantumGravity}
C.~Rovelli, {\em {Quantum Gravity}}.
\newblock Cambridge University Press, Cambridge, 2004.

\bibitem{ThiemannModernCanonicalQuantum}
T.~Thiemann, {\em {Modern Canonical Quantum General Relativity}}.
\newblock Cambridge University Press, Cambridge, 2007.

\bibitem{GreenBook1}
M.~B. Green, J.~H. Schwarz, and E.~Witten, {\em {Superstring Theory, Vol. 1:
  Introduction}}.
\newblock Cambridge University Press, Cambridge, 1988.

\bibitem{GreenBook2}
M.~B. Green, J.~H. Schwarz, and E.~Witten, {\em {Superstring Theory, Vol. 2:
  Loop Amplitudes, Anomalies and Phenomenology}}.
\newblock Cambridge University Press, Cambridge, 1988.

\bibitem{BTTVI}
N.~Bodendorfer, T.~Thiemann, and A.~Thurn, ``{Towards loop quantum supergravity
  (LQSG): I. Rarita-Schwinger sector},'' {\em Classical and Quantum Gravity}
  {\bf 30} (2013) 045006, {\tt arXiv:1105.3709 [gr-qc]}.

\bibitem{BaezAnInvitationTo}
J.~C. Baez and J.~Huerta, ``{An Invitation to Higher Gauge Theory},'' {\tt
  arXiv:1003.4485 [hep-th]}.

\bibitem{BaezInfiniteDImensionalRepresentations}
J.~C. Baez, A.~Baratin, L.~Freidel, and D.~K. Wise, ``{Infinite-Dimensional
  Representations of 2-Groups},'' {\tt arXiv:0812.4969 [math.QA]}.

\bibitem{Crane2CategorialPoincare}
L.~Crane and M.~D. Sheppeard, ``{2-categorical Poincare Representations and
  State Sum Applications},'' {\tt arXiv:math/0306440 [math.QA]}.

\bibitem{CraneMeasurableCategoriesAnd}
L.~Crane and D.~N. Yetter, ``{Measurable Categories and 2-Groups},'' {\em
  Applied Categorical Structures} {\bf 13} (2005) 501--516, {\tt
  arXiv:math/0305176 [math.QA]}.

\bibitem{YetterMeasurableCategories}
D.~N. Yetter, ``{Measurable Categories},'' {\em Applied Categorical Structures}
  {\bf 13} (2005) 469--500, {\tt arXiv:math/0309185 [math.CT]}.

\bibitem{BaezHigherGaugeTheory}
J.~C. Baez and U.~Schreiber, ``{Higher Gauge Theory},'' in {\em Categories in
  Algebra, Geometry and Mathematical Physics} ({A. Davydov et al}, ed.),
  (Providence, Rhode Island), pp.~7--30, AMS2007.
\newblock {\tt arXiv:math/0511710 [math.DG]}.

\bibitem{ThiemannQSD5}
T.~Thiemann, ``{Quantum spin dynamics (QSD) V: Quantum Gravity as the Natural
  Regulator of Matter Quantum Field Theories},'' {\em Classical and Quantum
  Gravity} {\bf 15} (1998) 1281--1314, {\tt arXiv:gr-qc/9705019}.

\bibitem{AshtekarRepresentationsOfThe}
A.~Ashtekar and C.~J. Isham, ``{Representations of the holonomy algebras of
  gravity and non-Abelian gauge theories},'' {\em Classical and Quantum
  Gravity} {\bf 9} (1992) 1433--1468, {\tt arXiv:hep-th/9202053}.

\bibitem{AshtekarRepresentationTheoryOf}
A.~Ashtekar and J.~Lewandowski, ``{Representation Theory of Analytic Holonomy
  C* Algebras},'' in {\em Knots and Quantum Gravity} (J.~Baez, ed.), (Oxford),
  Oxford University Press1994.
\newblock {\tt arXiv:gr-qc/9311010}.

\bibitem{NarnhoferCovariantQEDWithout}
N.~Narnhofer and W.~Thirring, ``{Covariant QED Without Indefinite Metric},''
  {\em Reviews in Mathematical Physics} {\bf 4} (1992) 197--211.

\bibitem{DiazHamiltonianFormulationOf}
A.~Diaz, ``{Hamiltonian formulation of eleven-dimensional supergravity},'' {\em
  Physical Review D} {\bf 33} (1986) 2801--2808.

\bibitem{NieuwenhuizenSupergravity}
P.~van Nieuwenhuizen, ``{Supergravity},'' {\em Physics Reports} {\bf 68} (1981)
  189--398.

\bibitem{HaagLocalQuantumPhysics}
R.~Haag, {\em {Local Quantum Physics: Fields, Particles, Algebras}}.
\newblock Springer, 2nd~ed., 1996.

\bibitem{ThiemannTheLQGString1}
T.~Thiemann, ``{The LQG string - loop quantum gravity quantization of string
  theory: I. Flat target space},'' {\em Classical and Quantum Gravity} {\bf 23}
  (2006) 1923--1970, {\tt arXiv:hep-th/0401172}.

\bibitem{CampbellBCH1}
J.~Campbell, ``{On a Law of Combination of Operators bearing on the Theory of
  Continuous Transformation Groups},'' {\em Proc. London Math. Soc.} {\bf
  s1-28} (1896) 381--390.

\bibitem{CampbellBCH2}
J.~Campbell, ``{On a Law of Combination of Operators (Second Paper)},'' {\em
  Proc. London Math. Soc.} {\bf s1-29} (1897) 14--32.

\bibitem{BakerBCH1}
H.~Baker, ``{Further Applications of Metrix Notation to Integration
  Problems},'' {\em Proc. London Math. Soc.} {\bf s1-34} (1901) 347--360.

\bibitem{BakerBCH2}
H.~Baker, ``{On the Integration of Linear Differential Equations},'' {\em Proc.
  London Math. Soc.} {\bf s1-35} (1902) 333--378.

\bibitem{BakerBCH3}
H.~Baker, ``{Alternants and Continuous Groups},'' {\em Proc. London Math. Soc.}
  {\bf s2-3} (1905) 24--47.

\bibitem{HausdorffBCH}
F.~Hausdorff, ``{Die symbolische Exponentialformel in der Gruppentheorie},''
  {\em Ber. Verh. Saechs. Akad. Wiss. Leipzig} {\bf 58} (1906) 19--48.

\bibitem{MagnusOnTheExponential}
W.~Magnus, ``{On the exponential solution of differential equations for a
  linear operator},'' {\em Comm. Pure Appl. Math.} {\bf 7} (1954) 649--673.

\bibitem{BratteliOperatorAlgebrasAnd1}
O.~Bratteli and D.~W. Robinson, {\em {Operator Algebras and Quantum Statistical
  Mechanics 1: C*- and W*-Algebras. Symmetry Groups. Decomposition of States}}.
\newblock Springer, 2nd~ed., 1987.

\bibitem{BratteliOperatorAlgebrasAnd2}
O.~Bratteli and D.~W. Robinson, {\em {Operator Algebras and Quantum Statistical
  Mechanics 2: Equilibrium States. Models in Quantum Statistical Mechanics}}.
\newblock Springer, 2nd~ed., 2003.

\bibitem{ThiemannQSD1}
T.~Thiemann, ``{Quantum spin dynamics (QSD)},'' {\em Classical and Quantum
  Gravity} {\bf 15} (1998) 839--873, {\tt arXiv:gr-qc/9606089}.

\bibitem{AshtekarPolymerAndFock}
A.~Ashtekar, J.~Lewandowski, and H.~Sahlmann, ``{Polymer and Fock
  representations for a scalar field},'' {\em Classical and Quantum Gravity}
  {\bf 20} (2003) L11--L21, {\tt arXiv:gr-qc/0211012}.

\bibitem{KaminskiBackgroundIndependentQuantizations1}
W.~Kaminski, J.~Lewandowski, and M.~Bobienski, ``{Background independent
  quantizations - the scalar field: I},'' {\em Classical and Quantum Gravity}
  {\bf 23} (2006) 2761--2770, {\tt arXiv:gr-qc/0508091}.

\bibitem{KaminskiBackgroundIndependentQuantizations2}
W.~Kaminski, J.~Lewandowski, and A.~Okol\'{o}w, ``{Background independent
  quantizations - the scalar field: II},'' {\em Classical and Quantum Gravity}
  {\bf 23} (2006) 5547--5585, {\tt arXiv:gr-qc/0604112}.

\end{thebibliography}\endgroup

\end{document}